# Anisotropy of exchange interactions in honeycomb ladder compound ReCl$_5$

A.A. Vorobyova,[1,2] A.I. Boltalin,[3] D.M. Tsymbarenko,[3] I.V. Morozov,[3] T.M. Vasilchikova,[1,4] V.V. Gapontsev,[5] K.A. Lyssenko,[3] S.V. Demishev,[6] A.V. Semeno,[7] S.V. Streltsov,[5] O.S. Volkova[1,4*]

[1]Functional Quantum Materials Laboratory, National University of Science and Technology "MISiS", Moscow 119049, Russia
[2]Higher School of Economics, Moscow 101000, Russia
[3]Department of Chemistry, M. V. Lomonosov Moscow State University, 119991 Moscow, Russian Federation
[4]Department of Physics, M. V. Lomonosov Moscow State University, 119991 Moscow, Russian Federation
[5]Institute of Metal Physics, RAS, Ekaterinburg 620990, Russia
[6]Institute for High Pressure Physics, RAS 142190, Russia
[7]Prokhorov General Physics Institute, RAS 119991, Russia
e-mail: os.volkova@yahoo.com

The Re$^{5+}$(5d$^2$) compounds possess large spin-orbital interaction which urges for large anisotropy, non – collinear structures and other phenomena. Here we present ReCl$_5$ composed by separate [Re$_2$Cl$_{10}$] units formed by edge – shared chlorine octahedra. It demonstrates the formation of antiferromagnetically ordered state in two steps at $T_{N1}$ = 35.5 K and $T_{N2}$ = 13.2 K seen in dc-, ac - magnetic susceptibility and in specific heat. At 4K it can be transformed to the state with spontaneous magnetic moment by relatively weak magnetic field $\mu_0 H$ = 0.5 T via metamagnetic phase transition. Ab initio calculations give anisotropic ferromagnetic exchange interactions $J1$ and $J2$ within and between rhenium pairs forming the zig-zag chains along the $a$ – axis. Pairs of zig-zag chains are coupled by ferromagnetic interaction $J3$ along the $c$ – axis into magnetic honeycomb ladders. The ladders are coupled by significantly weaker interaction $J4$.

## Introduction

The combination of strong electron correlation with strong spin-orbit and magnetic couplings make new materials perspective for spintronics [1,2]. Experimentally it can be found in 4$d$-5$d$ insulating compounds with partially filled $t_{2g}$ orbitals. Early transition metals pentachlorides $TM$Cl$_5$ ($TM$ = Nb, Ta, Mo, W, Re, Os) provide convenient experimental playground to study the interplay between the exchange and spin-orbit couplings in insulating regime with variation of number of $d$ electrons in isotypic structures. They are composed of isolated $TM_2$Cl$_{10}$ pairs loosely packed together via van der Waals forces. Each unit is formed by edge shared [$TM$Cl$_6$] octahedra with energetically favorable $t_{2g}$ group of orbitals with $e_g$ states lying much higher in energy, see e.g. [3]. The metal atoms are shifted from the octahedron centers away from each other which excludes direct overlap of metal orbitals. Insulating regime is supported by strongly ionic chemical bonds between metals and chlorine [4].

The dimers of $d^1$ metals are found and characterized in MoCl$_5$ and WCl$_5$ systems. Two modifications of molybdenum pentachloride α- MoCl$_5$ and ε- MoCl$_5$ with different stacking of chlorine atoms demonstrate ferromagnetic state formation below 22 and 14 K. Their saturated and effective magnetic moments are reduced to compare with $4d^1$ configuration ($S = 1/2$, $g = 2$) expected for Mo$^{5+}$ which is explained by the antiparallel orbital and spin moments. The intradimer and all interdimer exchange interactions were estimated as ferromagnetic. An anisotropy field was obtained theoretically and experimentally as 80 T and 12 T. These large values were explained by strong spin – orbit coupling. The WCl$_5$ compound possesses the same structure as α- MoCl$_5$ at room temperature. Theoretical calculations gave ferromagnetic intradimer and interdimer exchange interactions. However, this system ordered antiferromagnetically at low temperatures. The discrepancy was explained by structural transformation at 150 K [5].

The pairs of $d^2$ and $d^3$ metals reminding dimers can be found only for ReCl$_5$ and OsCl$_5$ which are $5d$ elements [6,7]. The pentachlorides of technetium and ruthenium do not exist. But $d^3$ electron configuration from the point of theory possesses completely quenched orbital degree of freedom and can be described by spin – only Hamiltonian with $S = 3/2$. The $d^2$ metal with $t_{2g}$ orbitals occupied in accordance with Hund's rule can demonstrate the orbitally dependent nature of the spin interactions. This in turn, may produce the significant biquadratic and triquadratic spin-spin interactions and different types of multipolar orders [8]. Here, we present a rare case of magnetism for $d^2$ system, organized by the pairs of rhenium. Available information about its properties was limited to the behavior of magnetic susceptibility at T > 77 which evidenced the reduced magnetic moment and large negative Weiss temperature $\Theta = -164$ K in the paramagnetic region [6]. In this work the ReCl$_5$ crystals were synthesized, the crystal structure at 100 K was clarified and measurements of thermodynamic properties down to 2K were performed. The quantum ground state of ReCl$_5$ was established experimentally and in first principles calculations.

**Materials and methods**

The ReCl$_5$ synthesis was carried out according to the following method. Rhenium metal has been preliminary purified of possible oxide impurities by heating in a hydrogen flow in a quartz reactor at 900°C for three hours. After cooling, the reactor with pure rhenium was vented with dried and purified chlorine, obtained by reacting hydrochloric acid with potassium permanganate. Then the reactor was heated up to 725°C. During the synthesis the substance was sublimated and transported in the gas flow into the cold zone of the reactor in the form of black crystals of rhenium(V) chloride. All further manipulations with ReCl$_5$ crystals were carried out in an argon atmosphere with an oxygen and water content of less than 0.1 ppm due to their high hygroscopicity and sensitivity to oxidation.

Single crystals were investigated by means of Bruker D8 Quest *X*-ray diffractometer (Photon III detector, microfocused Mo *K*α radiation tube) equipped with low temperature option down to 100 K. The absorption correction was performed using multiscan routine as implemented in SADABS. Crystal structure solution and refinement were performed using SHELX-2018 package [9]. Atomic positions were located using direct methods and refined using a combination of Fourier synthesis and least-squares refinement in isotropic and anisotropic approximations. Crystallographic parameters and final residuals for the single-crystal XRD experiment are given in Table S1 in Supplementary part. A summary of crystallographic data for the single-crystal experiment is available from CCDC 2356147.

The purity of the obtained sample was confirmed by PXRD data recorded using Tongda TD-3700 diffractometer (40kV, 30 mA, CuKα radiation, linear PSD detector) operated in Bragg-Brentano geometry (Fig. S1 of Supplementary part). Parameters refined using the Le-Bail method

are: sp.gr. $P2_1/c$, $a$ = 9.2048(6) Å, $b$ = 11.4324(10) Å, $c$ = 11.9244(11) Å, $\beta$ = 109.310(5)°, $V$ = 1184.25(18) Å$^3$ ($R$ = 0.0136, $wR$ = 0.0173).

Thermodynamic measurements, i.e. magnetization, ac - susceptibility and specific heat, were measured by Heat Capacity, AC Magnetic System and VSM Magnetometer options of Physical Properties Measurement System PPMS 9T "Quantum design". The sample is hygroscopic. All manipulations with it were performed in a glovebox to avoid its contact with air. Electron spin resonance (ESR) study was carried out using an *X*-band ESR spectrometer CMS 8400 (ADANI) ($f$ = 9.4 GHz, $\mu_0 H \leq 0.7$ T) equipped with a low temperature mount, operating in the range T = 6-300 K. The effective g-factor has been calculated with respect to a BDPA (a,g -bisdiphenyline-b-phenylallyl) reference sample with $g_{et}$=2.00359. High frequency (60 GHz) magnetic resonance at helium temperatures was studied with the help of the ESR spectrometer designed at Prokhorov General Physics Institute of Russian Academy of Sciences equipped with 8 T superconducting magnet and utilizing cylindrical cavity tuned to $TE_{011}$ mode. With this spectrometer direct absorption P(H) was measured.

The first principles calculations were performed using the Vienna Ab-initio Simulation Package (VASP). We utilized the generalized gradient approximation (GGA) and the projector augmented-wave (PAW) method [10]. In order to take into account strong electron-electron correlations we employed the GGA+U approach [11]. The on-site Coulomb repulsion parameter $U$ and intra-atomic Hund's exchange $J_H$ for Re were chosen to be 2 eV and 0.5 eV, respectively close to available literature data [12-14]. The cut-off energy was chosen to be 500 eV and the 2x3x3 k-mesh, centered at the Γ point, was used for integration over the Brillouin zone. The Wigner-Seitz radii equal to 1.376 Å for Re and 1.005 Å for Cl. The self-consistency threshold was chosen to be $10^{-7}$ eV. The exchange integrals were calculated by the 4-configurations method [15].

**Results**
**Crystal structure**

The crystal structure of rhenium(V) chloride $ReCl_5$ consists of dimeric molecules $[Re_2Cl_{10}]$ with two crystallographically independent rhenium atoms. Each dimer is formed by edge shared octahedra as shown in left panel of Fig.S2 in Supplementary part. The Re-Cl distances vary in the range from 2.229(2) till 2.475(2) Å at 100K. The Re(1)-Re(2) distance within dimer amounts 3.734(1) Å. They are shifted from the octahedron centers away from each other along $a$ – axis. The shortest interdimer distances between rhenium atoms are 5.778(1) Å and 5.776(1) Å along $a$ – and $c$ – axes as shown in Fig.1. The basal planes of dimeric molecules in neighboring ladders are mutually orthogonal as is shown in right panel of Fig.S2 in Supplementary part.

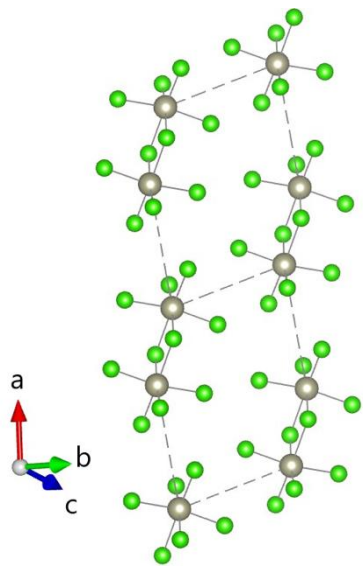

Fig. 1. The set of dimerized [Re$_2$Cl$_{10}$] units along a – axis. Rhenium atoms are shown by large spheres, chlorine atoms are small spheres. Thin dotted lines are the shortest interdimer Re – Re distances.

### Specific heat and magnetization

The temperature dependence of reduced specific heat $C_p/T$, shown in Fig. 2, demonstrates two smeared maxima at $T_{N1}$ = 35.5 K and $T_{N2}$ = 13.2 K. The lattice contribution $C_{lat}$, shown by solid line in the inset to Fig. 2, was approximated by the sum of Debye [16] and Einstein [17] functions, with weights of $a_D$ = 3.8 with $\Theta_D$ = 370 K and $a_E$ = 2.2 with $\Theta_E$ = 100 K. Total magnetic entropy $S_m$ is shown in Fig. S3 in Supplementary section. It is obtained by integration of $C_m/T$ function calculated by subtraction of $C_{lat}$ from $C_p$ shown in the inset to Fig. S3 in Supplementary section. Magnetic entropy $S_m$ reaches 13.4 J/mol K indicating that the ground state is characterized by a (nearly) degenerate quintet. Indeed, in both *LS*- or *jj*-coupling scheme for $t^2_{2g}$ configuration with the effective orbital moment $L$=1 and total spin $S$=1 has effective total moment $J$=2 [18]. Therefore, the magnetic entropy can be estimated as $S_m = R\ln(2J+1) = R\ln5 = 13.4$ J/molK. It is worth mentioning that the ReCl$_6$ octahedra are distorted (point group is $C_1$) and therefore degeneracy of the quintet must be completely lifted (except Kramer's degeneracy), but our analysis shows that corresponding splittings are not large enough since magnetic contribution to the entropy saturates at $R\ln5$ at temperatures ~50K.

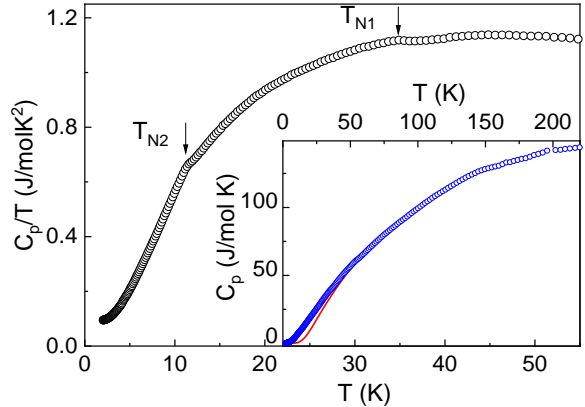

Fig. 2. Reduced specific heat $C_p/T$ of ReCl$_5$. Arrows mark the temperatures of magnetic ordering $T_{N1}$ and $T_{N2}$. The inset represents the specific heat $C_p$ and lattice contribution.

The temperature dependence of magnetic susceptibility $\chi_{dc}(T)$ of ReCl$_5$, shown in Fig. 3, obeys Curie – Weiss law at high temperatures, i.e. $\chi_{dc} = \chi_0 + C/(T-\Theta)$. The parameters of the fit in the range 200 -300 K are $\chi_0 = 3.15 \cdot 10^{-4}$ emu/mol, $C$ = 0.26 emuK/mol, $\Theta$ = -61 K. The effective moment $\mu_{eff}=\sqrt{8C}$ equals to 1.44 $\mu_B$/f.u. This value agrees reasonably with other Re$^{5+}$ or Os$^{6+}$ compounds which demonstrate 1.6 – 2.0 $\mu_B$ per 5$d^2$ ion [1,2]. Negative Weiss temperature is an attribute of antiferromagnetically ordered state below $T_{N1}$. Low frustration ratio $\Theta/T_{N1}$ < 2 indicates for the presence of hidden ferromagnetic component in exchange interactions. At low temperature $\chi(T)$ shows a kink at $T_{N1}$ = 35.5 K and a narrow peak at $T_{N2}$ = 13.2 K, as shown in Fig. 3.

The *ac* – magnetic susceptibility consists mainly from the real part $\chi'$. As shown in the inset to Fig. 3, both $\chi_{ac}$ and $\chi'$ show peaks at $T_{N1}$ and $T_{N2}$ corresponding the antiferromagnetic state formation. Absence of imaginary part $\chi''$ excludes spin dynamics observed in the systems with metallic conductivity [19].

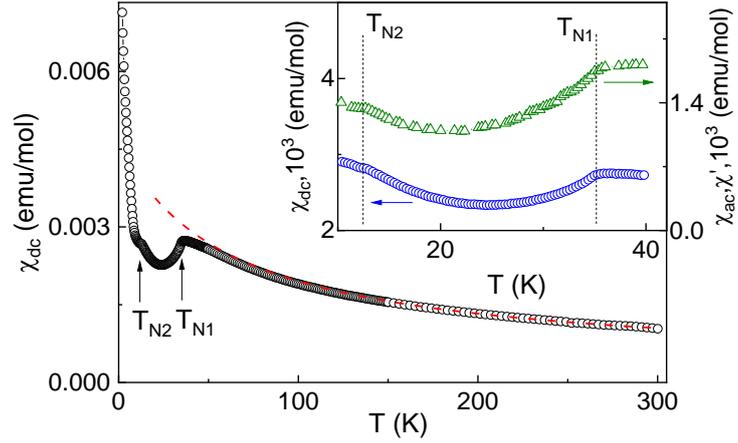

Fig. 3. Magnetic susceptibility $\chi_{dc}$ of ReCl$_5$ measured after cooling at $\mu_0H = 0.1$ T. Dotted line is a fit with Curie – Weiss law. Inset shows magnetic susceptibility $\chi_{dc}$ measured at $\mu_0H = 0.1$ T after cooling without field(circles) and ac – susceptibility $\chi_{ac}$ with its real part $\chi'$ measured at 5kHz (triangles). Arrows and vertical dashed lines mark $T_{N1}$ and $T_{N2}$.

Typical field dependence of magnetization of ReCl$_5$ is shown in Fig. 4. All curves measured at $T < T_{N2}$ demonstrate a kink at $\mu_0H = 0.5$ T seen more clearly in the derivative which corresponds the metamagnetic phase transition. This indicates for Ising type anisotropy in ReCl$_5$. The low value of magnetization at 9 T can be attributed to the subtraction of the orbital angular momentum from the spin only value, M = gS-L. The saturated moment can be achieved in a field stronger than the anisotropy and magnetic exchange interactions presented in the system which is much higher than experimentally accessible magnetic fields.

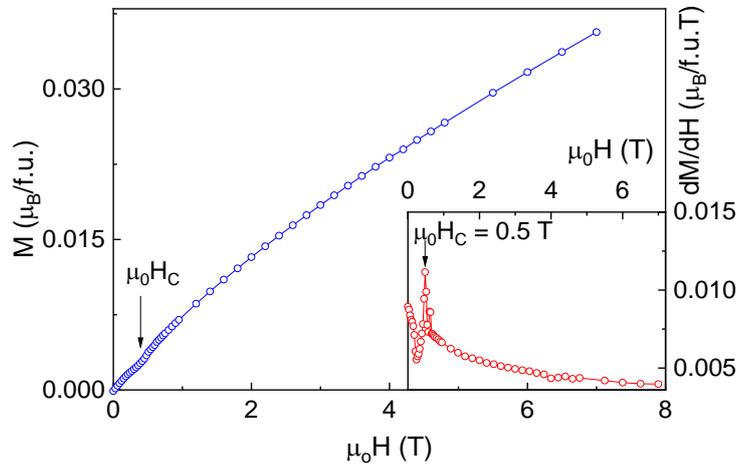

Fig. 4. Magnetization curve (main panel) and its derivative (inset) taken at T = 4 K for ReCl$_5$. The arrow indicates the field of metamagnetic phase transition.

**ESR spectroscopy**

The ESR data for ReCl$_5$ at 9.4 GHz and 60 GHz are shown in Fig. 5. The temperature dependence (Fig. 5a) demonstrates typical paramagnetic behavior for some paramagnetic center with g-factor $g = 1.97\pm0.03$, which has hyperfine splitting. The hyperfine structure can be associated with the interaction of unpaired electron and nuclear magnetic moments of the isotopes

$^{185}$Re ($I = 5/2$, 37.4% natural abundance) and $^{187}$Re ($I = 5/2$, 62.6% natural abundance). Since the difference between the magnetic moments of these two nuclei is about 1% (3.1871 μ/μ$_N$ and 3.2197 μ/μ$_N$ for $^{185}$Re and $^{187}$Re, respectively), it cannot be resolved.

The high-frequency (60 GHz) ESR spectrum $P(H)$ at liquid helium temperature is formed by an asymmetric broad line, which can be de described by the sum of three Lorentz lines with g-factors $g_1$=2.24±0.01, $g_2$=2.035±0.001 and $g_3$=1.88±0.04 (Fig. 5b). The integrated intensities for the lines with g-factors $g_1$ and $g_2$ are comparable, whereas the line with $g_3$ is significantly weaker. In this case, the hyperfine structure is not resolved, presumably due to the enhancement of the relaxation parameter in this spectral range.

Since the ESR line at 9.4 GHz with the same position is observed above and below antiferromagnetic transition at $T_{N1}$~35 K, it is difficult to expect that observed ESR modes are caused by Re$^{5+}$ (d$^2$) ions having quintet ground state with $J = 2$. Moreover, this type of magnetic resonance was not reported for rhenium Re$^{5+}$ ions in any compound so far. Usually, Re ions demonstrate ESR with the effective spin multiple to one half, i.e in the d$^1$, d$^3$ or d$^5$ configurations. For Re$^{6+}$ (d$^1$) the g-factors of the spin Hamiltonian in various compounds measured in different frequency ranges are within the limits 1.61-1.82 and can be characterized by moderate anisotropy [20, 21]. The rhenium ion Re$^{2+}$ (d$^5$) cannot be present in this compound due to the conditions of the synthesis. However, it is likely that oxidation of rhenium metal with chlorine vapor resulted in the formation of rhenium pentachloride with a small number of defective, underoxidized rhenium atoms Re$^{4+}$ (d$^3$) surrounded by less than six chlorine atoms in two crystallographic positions. That is, the ReCl$_5$ powder contains an extremely small number of Re$^{4+}$ - Re$^{5+}$ dimers. These dimers at low temperatures form unit with effective spin $S = 5/2$ due to strong ferromagnetic intra-dimer exchange interaction. Rhenium defective dimers produce absorption lines with g-factors 1.88, 2.03 and 2.24 similar to (Ph$_4$As)$_3$[Re$_2$OCl$_{10}$] [22].

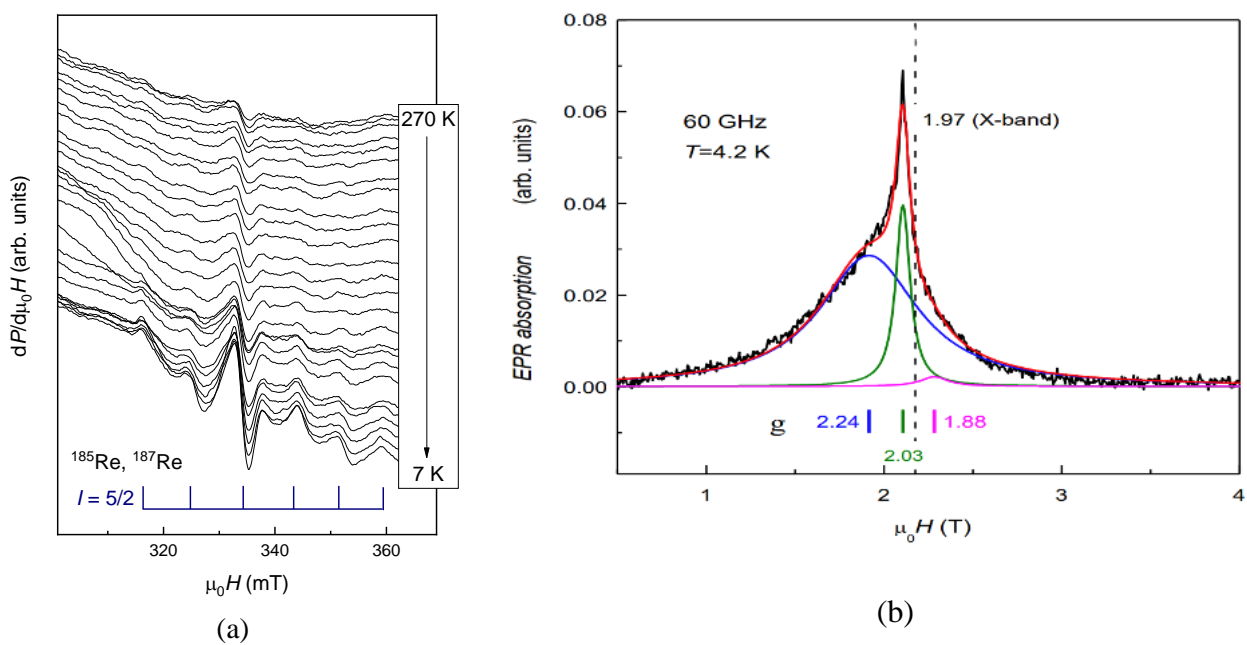

Fig. 5. (a) The evolution of the 9.4 GHz ESR spectra d$P$/dμ$_0H$ with temperature for ReCl$_5$. (b) High frequency (60 GHz) $P$(μ$_0H$) ESR spectrum at T=4.2 K. Three Lorentzian components with various g

factors are shown by solid lines. The dashed line corresponds to g=1.97 following from the X band experiment.

The analysis of magnetic and thermodynamic properties presented in the previous sections clearly shows that $Re^{5+}$ ($d^2$) ions have ground state with $J=2$, in contrast to the ESR data. We believe that this discrepancy has a natural explanation. The $d^2$ states with $J=2$ apparently correspond to the bulk material, whereas the $d^5$ states visible by ESR method are due to defective dimers, making a minor contribution to the magnetization and specific heat.

**First principles calculations**

In order to elucidate effects of the crystal-field splitting, spin-orbit coupling and strong Hubbard correlations we performed nonmagnetic GGA and magnetic GGA+SOC, GGA+$U$, GGA+$U$+SOC calculations, see Figs. 6-8. It is tempting to call the Re-Re pair dimers, which typically manifest themselves via formation of the bonding and antibonding states. However, nonmagnetic GGA band structure presented in Fig. 6 clearly demonstrates that there is no any bonding – antibonding splitting typical for dimerized systems. In fact, the shortest Re-Re distance in $ReCl_5$, 3.73 Å, is much larger than the metal-metal distance in metallic Re (2.74 Å [23]). Therefore, it is not surprising that no physics related to the structural Re-Re "dimers" (which we prefer to call pairs) is observed. In nonmagnetic GGA band structure (see Fig. 6) one can observe a bunch of nearly unsplit $t_{2g}$ bands (24 bands since there are 8 Re atoms in the unit cell; from -0.5 to 0.5 eV) very well isolated from the $e_g$ bands (16 bands from 2.5 to 3.7 eV) by the crystal field splitting of about 3.1 eV.

Next, one can clearly see that strong electronic correlations are unable to open the band gap and $ReCl_5$ turns out to be half-metallic in the GGA+$U$ approximation (top panel of Fig. 7). Moreover, while the spin-orbit coupling is large for heavy elements, it does not result in the insulating ground state alone as well, see middle panel of Fig. 7. Thus, it is crucial to have both contributions, in GGA+$U$+SOC $ReCl_5$ finally becomes insulating with the band gap 0.2 eV (bottom panel of Fig. 7). Therefore, this material belongs to the class of the spin-orbit assisted Mott insulator similar to e.g. famous $Sr_2IrO_4$ [24].

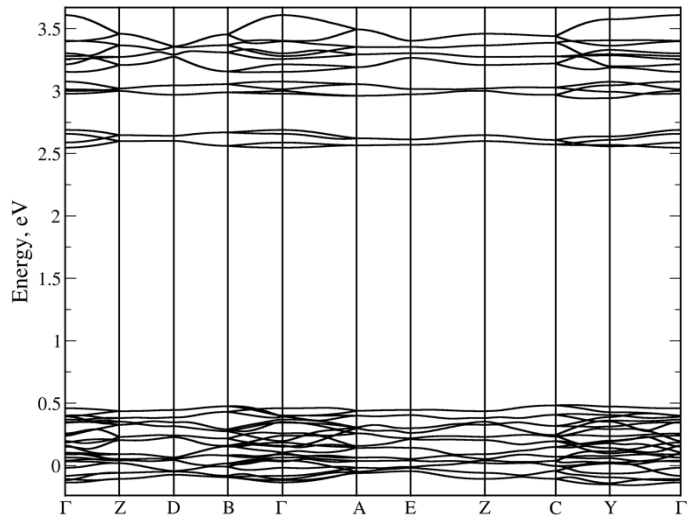

Fig. 6. Band structure for ReCl$_5$ obtained within the nonmagnetic GGA approximation. The Fermi level corresponds to zero energy.

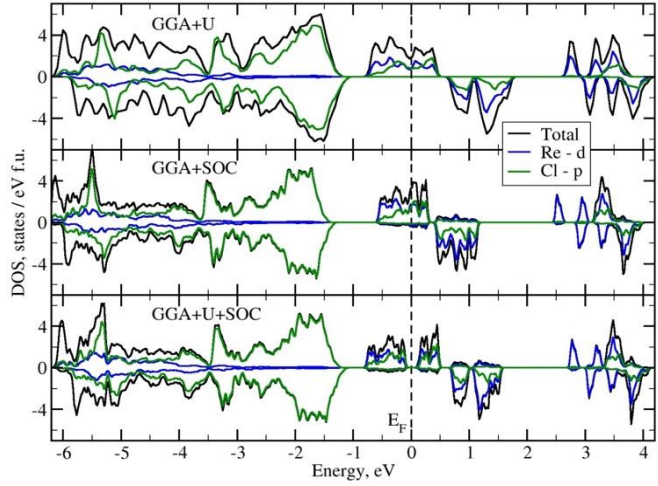

Fig. 7. Total and partial density of states. The top panel shows results calculated using GGA+$U$. For the plot in the middle panel we use GGA taking into account spin-orbit coupling (SOC). In the bottom panel we applied both SOC and Hubbard $U$ (GGA+$U$+SOC). The Fermi level is at 0 eV, positive (negative) DOS corresponds to the spin majority (minority).

The band gap is formed mostly by the spin-orbit split atomic $t_{2g}$ states with strong admixture of Cl-$p$ orbitals. The splitting between different spin-subbands is ~1.2 eV. The $e_g$ states of Re are placed much higher in energy, at ~3-4 eV above the Fermi level.

The spin-orbit coupling not only qualitatively changes electronic structure of ReCl$_5$, but also modifies substantially the magnetic moments and the exchange coupling as it will be shown below. The $t_{2g}$ orbitals can be described by the effective orbital momentum $l$=1 [23], so that the spin-orbit coupling can be written as $\hat{H}_{SOC} = -\lambda ls$ (neglecting off-diagonal $t_{2g}$ - $e_g$ matrix elements). On the one hand, this leads to inversion of the multiplets [23], so that in spite of the less-than-half-filling situation the lowest in energy is a state with the total effective momentum $J = L+S$. This also must result in reduction of the magnetic moment $M=gS-L$ [25] and this is exactly what is seen in the GGA+$U$+SOC calculations: the absolute value of the spin moment was found to be 1.41$\mu_B$ with components of (-0.468; -0.617; 1.181)$\mu_B$, while the orbital contribution is directed in (nearly) opposite direction being (0.176; 0.279; -0.444) $\mu_B$.

Strong spin-orbit coupling induces the anisotropy of the exchange interaction, which e.g. plays a key role in famous Kitaev materials [26-30]. Therefore, we calculated not only isotropic (Heisenberg) exchanges, but all elements of the exchange tensor for the Heisenberg model written as

$$H = 1/2 \sum_{i \neq j} \vec{S_i} J_{ij} \vec{S_j}, \quad (1)$$

where $J_{ij}$ is the matrix for each pair of site indexes $i$ and $j$. The results (in meV) are given below:

$$J_1 = \begin{pmatrix} -15.2 & 12.1 & -3.9 \\ 10.1 & -17.5 & -0.1 \\ -4.5 & 0.1 & -6.9 \end{pmatrix}, J_2 = \begin{pmatrix} -16.0 & 14.9 & 3.2 \\ 13.4 & 0.2 & -1.9 \\ 3.9 & -2.0 & -15.9 \end{pmatrix},$$

$$J_3 = \begin{pmatrix} -31.5 & -0.7 & 2.1 \\ 0.2 & -25.5 & -0.1 \\ 3.1 & 0.0 & -0.7 \end{pmatrix}, J_4 = \begin{pmatrix} -4.3 & 1.4 & -1.6 \\ 5.4 & -2.4 & -1.5 \\ 2.1 & 0.5 & -13.4 \end{pmatrix}.$$

Three nearest neighbor Re-Re pairs were considered in these calculations. $J1$ corresponds to the shortest Re-Re distance. Together with $J2$ they describe the exchange interaction within alternating Re chains running (nearly) along the *a*-axis. Surprisingly, the strongest exchange interaction was observed not in the chains, but across the legs of "honeycomb ladders", corresponding to exchange $J3$, see Fig. 3c. We used the same coordinate system for all Re-Re pairs, shown in inset of Fig. 8, where *z*-axis is perpendicular to the plane defined by the common edge of two $ReCl_6$ octahedra constituting the pair and the Re-Re bond by itself. The *x*- and y- axes are approximately directed to the in-plane ligands (this choice of coordinate systems is typical for Kitaev materials, see e.g. [31]).

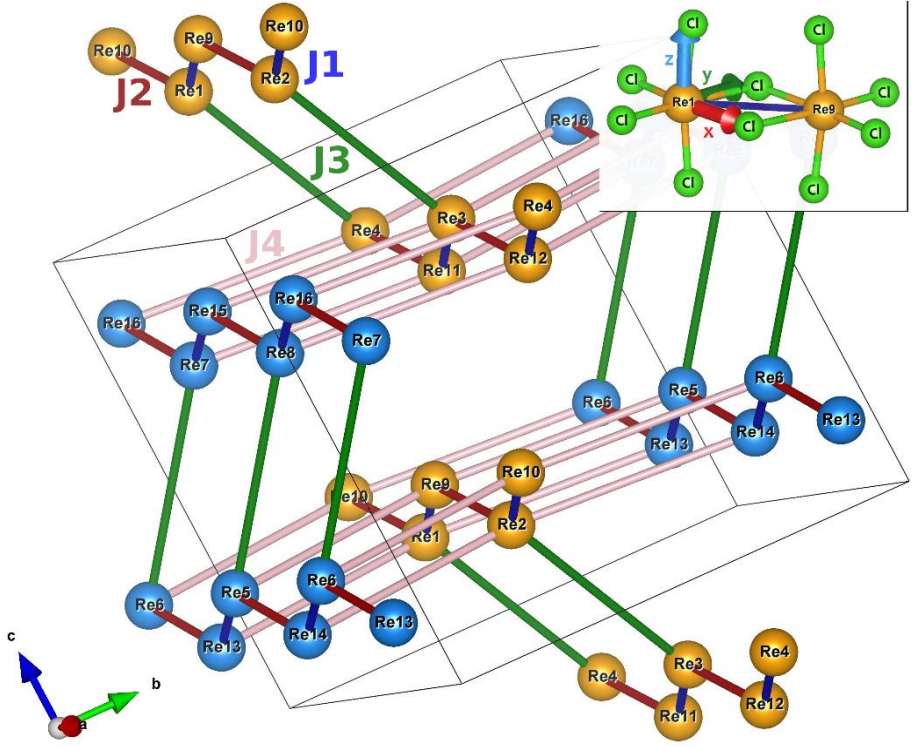

Fig. 8. The main exchange interaction paths in $ReCl_5$: $J1$ (intrapair) and $J2$, $J3$ (interpair), forming magnetic "honeycomb" chains along *a* – axis, and $J4$. The spheres of same color show co-directed magnetic moments.

We see that the exchange interaction is indeed strongly anisotropic, but the antisymmetric contribution (corresponding to Dzyaloshinskii-Moryia interaction) is not generally large (except $J4$); It has to be mentioned that the nearest honeycomb ladders running along *a*-axis are nearly orthogonal to each other. This strongly reminds the situation with the famous Shastry-Sutherland lattice and makes $ReCl_5$ especially interesting for further studies.

While exchange tensors contain both positive and negative terms, one can clearly see that ferromagnetic (negative) contributions dominate. This is seemingly inconsistent with a global antiferromagnetism following from the negative experimental Curie-Weiss temperature. Detailed simulations of magnetic properties of such a complex spin system using e.g. Quantum Monte-Carlo method is far beyond the scope of the present work, but one rather important point has to be mentioned already at this stage. It has been recently noted that interplay of a large spin-orbit coupling and the on-site Coulomb interaction often result in reach multiplet structure in late transition metal compounds and low-lying multiplets can substantially affect temperature dependence of the magnetic susceptibility [32-34]. Such Van-Vleck-like (temperature dependent) contributions are known from long-ago [35,36].

The extreme situation presented in the double perovskites $Ba_2AOsO_6$, where A is Zn, Mg or Ca and Os with electronic $d^2$ configuration, is similar to our case in $ReCl_5$. In cubic symmetry, Os ions should have a vanishing magnetic moment (but the octupolar moment remains) in the ground state [37]. The high-temperature tail of its inverse magnetic susceptibility shows large AFM Curie-Weiss temperature [34]. Although the local symmetry of Re ions in $ReCl_5$ is much lower than cubic (and therefore octupolar state cannot be realized), the same physics of Van-Vleck-like contributions can be relevant in our situation and may lead to the AFM Curie-Weiss temperature. Further studies including true many-body calculations beyond DFT level have to be carried out to check this scenario.

**Discussion**

The $ReCl_5$ composed of $Re^{5+}$ ions with two d-electrons possesses spin and effective orbital moments equal to $S = 1$ and $L = 1$ producing the $J = 2$ ground state as seen from our analysis of the magnetic entropy, which saturates at $S_m = R\ln5$ at low temperatures. Partially suppressed magnetic moments order antiferromagnetically below $T_{N1} = 35.5$ K and $T_{N2} = 13.2$ K. Ground state can be easily influenced by a rather weak magnetic field $\mu_0H = 0.5$ T and modified into the phase with spontaneous moment via metamagnetic phase transition.

In paramagnetic region $ReCl_5$ demonstrates small effective magnetic moment $\mu_{eff}= 1.44$ $\mu_B$/f.u and negative Weiss temperature $\Theta = -61$ K while theoretical calculations give ferromagnetic sign for all exchange interactions. This can be due to the influence of orbital subsystem observed earlier in other 5d elements compounds, e.g. double perovskites $Ba_2NaOsO_6$, $Ba_2MgReO_6$ and $Ba_2ZnReO_6$. They are ferromagnets with negative $\Theta$ [38,39]. Theoretical analysis of deviations from Curie – Weiss law in $TMCl_5$ family similar to that done for double perovskites [40] will be useful to unveil the role of spin – orbit coupling for the formation of quantum ground state in the geometry with edge – shared octahedra fulfilled with 5d atoms.

The obtained $X$ band (9.4 GHz) and $V$ band (60 GHz) electron spin resonance data suggest that ESR signal may be caused by defect paramagnetic centers located at $ReCl_5$ surface corresponding mainly to $Re^{2+}$ ($d^5$) states with possible minor contribution of $Re^{4+}$ - $Re^{5+}$ dimers.

DFT calculations clearly demonstrate that while there are structural Re-Re dimers, there is no even a hint on formation of bonding-antibonding splitting in the electronic band structure. Instead, these are the spin-orbit coupling and strong electronic correlations which define both magnetic and electronic properties of this material. They conspire to give the insulating ground state so that $ReCl_5$ must be considered as a spin-orbit assisted Mott insulator. The spin-orbit coupling also partially suppresses magnetic moments and strongly affects the exchange interaction making it anisotropic. Probably, competition between anisotropies of main exchange interactions results in two – step formation of magnetic structure due to the appearance of incommensurate and commensurate structures below $T_{N1}$ and $T_{N2}$ which is necessary to be checked with neutron scattering. The exchange coupling is ferromagnetic within the honeycomb ladders, but expected to be antiferromagnetic between them. It is the first case of observation the localized magnetic moment on $Re^{5+}$ in the system with Re-Cl-Re direct bonds.

**Conclusion**

The $ReCl_5$ is the first example of closely packed rhenium atoms in the spin-orbit assisted Mott insulator regime. The architecture of strongly anisotropic magnetic interactions highlights fashion honeycomb ladders within its structure. Thermodynamic properties allow to attribute it to an antiferromagnet with non – trivial sequence of two magnetic phase transitions and reduced magnetic moment.

**Acknowledgments**

V.V.G thanks the Russian Ministry of Science and Education for support of analysis of the electronic structure of $ReCl_5$ via the "Quantum" program (No 122021000038-7. Calculations of exchange constants were supported by the Russian Science Foundation though the RSCF 23-12-00159 project. V.T.M. thank RSCF grant 22-42-08002 for resonant measurements. V.A.A. thanks RSCF 22-72-10034 project for the synthesis. M.I.V. thanks RSCF project 22-43-02020 for X-Ray analysis. V.O.S. thanks Ministry of Science and Higher Education of the Russian Federation within the framework of the Priority-2030 strategic academic leadership program at NUST MISIS for thermodynamic measurements.

**Author contributions**

**A.A. Vorobyova:** Investigation, Resources **A.I. Boltalin:** Formal analysis, Investigation **D.M. Tsymbarenko:** Investigation, Methodology **I.V. Morozov:** Formal analysis, Visualization **T.M. Vasilchikova:** Investigation, Methodology **V.V. Gapontsev:** Investigation, Methodology **K.A. Lyssenko:** Investigation, Funding acquisition, Supervision **S.V. Demishev**: Investigation, Funding acquisition Writing - original draft **A.V. Semeno:** Investigation, Resources **S.V. Streltsov:** Conceptualization, Writing - original draft, Funding acquisition **O.S. Volkova:** Funding acquisition, Validation, Writing - review & editing

**Declarations**
**Conflict of interests** The authors declare that they have no conflict of interest.

**Anisotropy of exchange interactions in honeycomb ladder compound ReCl$_5$**

A.A. Vorobyova,[1,2] A.I. Boltalin,[3] D.M. Tsymbarenko,[3] I.V. Morozov,[3] T.M. Vasilchikova,[1,4] V.V. Gapontsev,[5] K.A. Lyssenko,[3] S.V. Streltsov,[5] S.V. Demishev,[6] A.V. Semeno,[7] O.S. Volkova[1,4*]

[1]Functional Quantum Materials Laboratory, National University of Science and Technology "MISiS", Moscow 119049, Russia
[2]Higher School of Economics, Moscow 101000, Russia
[3]Department of Chemistry, M. V. Lomonosov Moscow State University, 119991 Moscow, Russian Federation
[4]Department of Physics, M. V. Lomonosov Moscow State University, 119991 Moscow, Russian Federation
[5]Institute of Metal Physics, RAS, Ekaterinburg 620990, Russia
[6]Institute for High Pressure Physics, RAS 142190, Russia
[7]Prokhorov General Physics Institute, RAS 119991, Russia
e-mail: os.volkova@yahoo.com

Table S1 Crystal lattice parameters of ReCl$_5$ at 100K.

| Compound | Rhenium(V) chloride |
|---|---|
| formula | Re$_2$Cl$_{10}$ |
| M$_r$ | 726.90 |
| T, K | 100(2) |
| crystal system | monoclinic |
| space group | $P\,2_1/c$ |
| $a$ [Å] | 9.1796(11) |
| $b$ [Å] | 11.3995(13) |
| $c$ [Å] | 11.9243(13) |
| β [°] | 108.955(3) |
| $V$ [Å$^3$] | 1180.1(2) |
| Z | 4 |
| Dc [g·cm$^{-3}$] | 4.091 |
| refls collected / Rint | 15872 / 0.0652 |
| parameters | 109 |
| $R_1(l \geq 2\sigma(I)$ / $wR_2$(all) | 0.0389 / 0.0672 |
| CCDC | 2356147 |

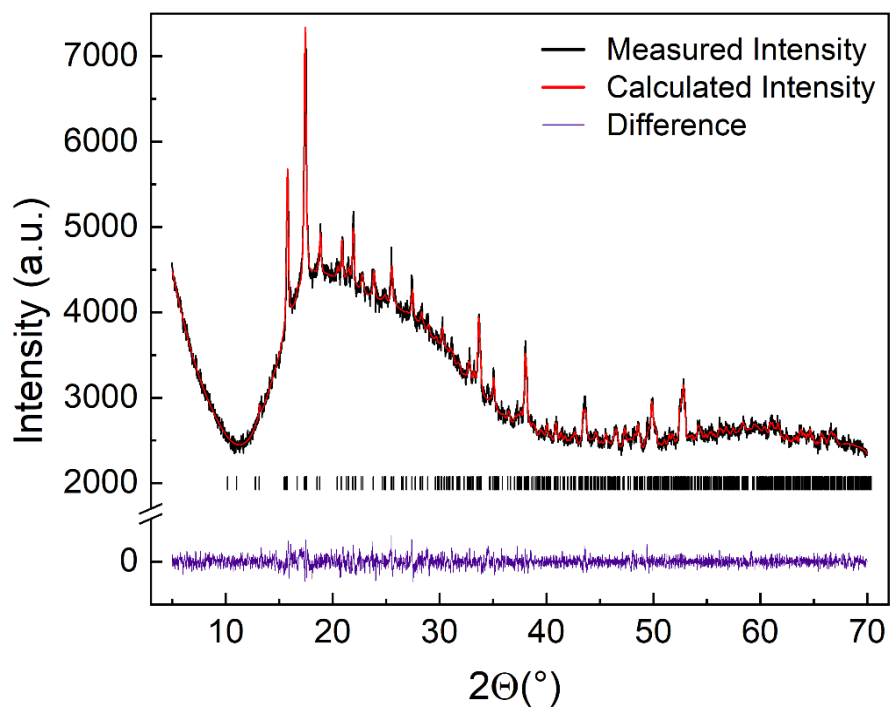

Fig.S1. Powder X-ray pattern of ReCl$_5$ fitted by the Le-Bail method at T = 300K.

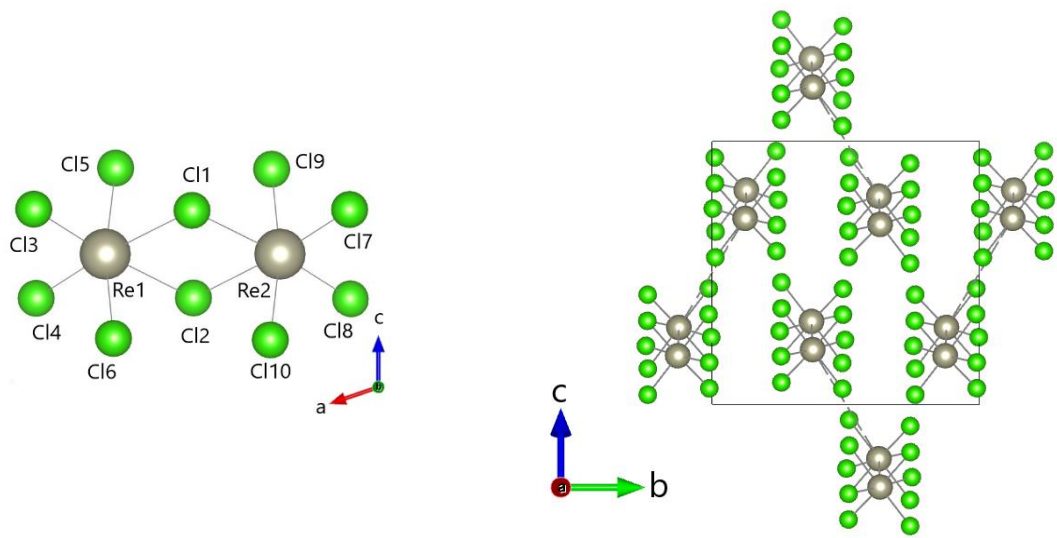

Fig. S2. Left: dimeric molecule [Re$_2$Cl$_{10}$]. Right: the projection of ReCl$_5$ in the bc-plane. Thin dotted lines are the shortest interdimer Re – Re and distances.

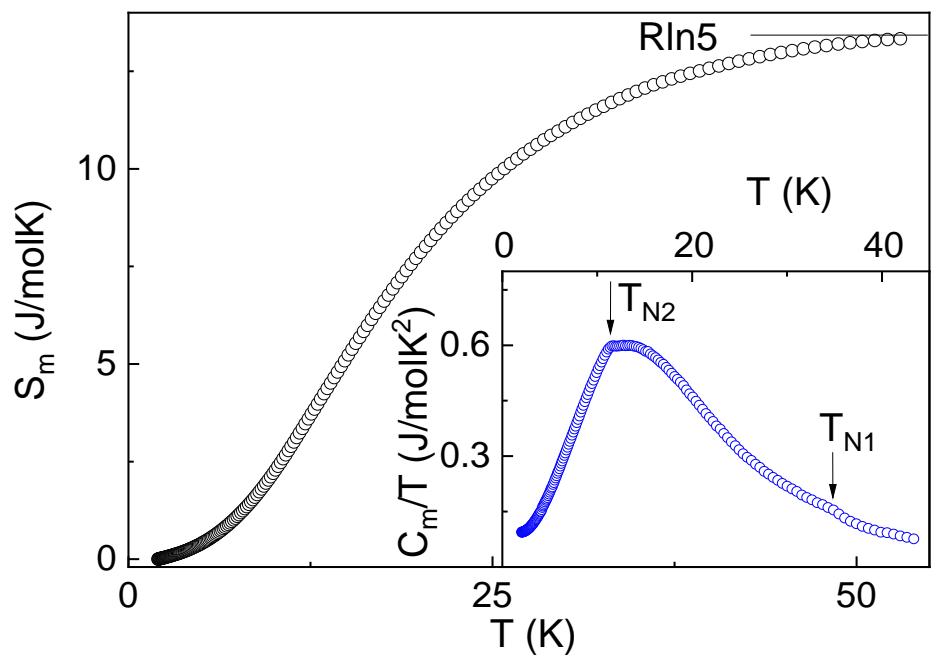

Fig. S3. Temperature dependencies of magnetic entropy $S_m$ (main panel) and magnetic specific heat divided by temperature $C_m/T$ (inset) of ReCl$_5$. Horizontal stick is a theoretical value of $S_m$. Arrows indicate the temperatures of magnetic phase transitions.